\documentclass[pra,aps,twocolumn,amsmath,amssymb]{revtex4}
\usepackage{graphicx}
\usepackage{dcolumn}
\usepackage{bm}
\usepackage[extension=xxx]{hyperref}

\newcommand{\beq}{\begin{equation}}
\newcommand{\eeq}{\end{equation}}
\newcommand{\beqa}{\begin{eqnarray}}
\newcommand{\eeqa}{\end{eqnarray}}

\def\beq{\begin{equation}}

\usepackage{color}
\begin{document}
\title{Phase space manipulations of many-body wavefunctions}
\author{G. Condon}
\author{A. Fortun}
\author{J. Billy}
\author{D. Gu\'ery-Odelin}

\affiliation{Universit\'e de Toulouse ; UPS ; Laboratoire Collisions Agr\'egats R\'eactivit\'e, IRSAMC ; F-31062 Toulouse, France} 
\affiliation{CNRS ; UMR 5589 ; F-31062 Toulouse, France}

\begin{abstract}
We explore the manipulation in phase space of many-body wavefunctions that exhibit self-similar dynamics, under the application of sudden force and/or in the presence of a constant acceleration field. For this purpose, we work out a common theoretical framework based on the Wigner function. We discuss squeezing in position space, phase space rotation and its implications in cooling for both non-interacting and interacting gases, and time reversal operation. We discuss various optical analogies and calculate the role of spherical-like aberration in cooling protocols. We also present the equivalent of a spin-echo technique to improve the robustness of velocity dispersion reduction protocols.
\end{abstract}

\maketitle

Phase space manipulations are at the heart of astonishing developments in atomic and molecular physics. Laser cooling of cold atom samples provides a spectacular example with the increase of the phase space density by populating a large number of photon modes through the dissipative mechanism provided by spontaneous emission \cite{CCTDGO}. Alternatively, evaporative cooling exploits the irreversible nature of 3D elastic collisions to increase the phase space density of a sub-ensemble of confined particles. The phase space density of atomic beams has also been increased with similar techniques \cite{MESCHEDE,DGO2005}. The demonstration of Maxwell's Demon devices that combined conservative potentials with an irreversible step belongs to the same kind of phase space manipulations \cite{RAIZENSCIENCE,muga06}. In the absence of dissipative mechanisms, the phase space volume is conserved. The manipulations that can be carried out in phase space with well engineered time-dependent conservative potentials involve separately or in combination: translation (used for instance in the slowing down of atomic or molecular packets \cite{Crompvoets,Carr,Reinaudi,Raizen07,RaizenPRL}) and deformation \cite{Carr,Morinaga,DGO2013} including  compressions either in position or momentum space \cite{Chu,DKick97,Marechal,Myrskog,Morinaga,Aoki,Goldberg} or magnification \cite{Murthy,Campo}. Such methods are quite general, they do not rely on a specific internal structure and can therefore be applied to a large class of particles including neutrons \cite{SNS86,SNR86}.

In Ref.~\cite{DKick97}, the authors proposed a phase space manipulation for velocity dispersion reduction of a non-interacting wave packet based on phase imprinting. This method has proven to cool very efficiently thermal and Bose-condensed atomic samples, leading recently to temperatures as low as 50 pK~\cite{Kovachy}. This efficient narrowing of the velocity dispersion is of great interest in metrology measurements based on atom interferometry~\cite{Muntinga,McDonald} and also for realizing quantum simulations~\cite{Jendrzejewski}.


Only very recently, such techniques have started to be applied to strongly interacting atoms~\cite{Murthy}. In this paper, we precisely investigate the generalization of such a cooling concept for manipulating in phase space many-body quantum systems that exhibit self-similar dynamics \cite{adolfo,a13,a38,a39,a14,a16,a10,a41,a42,a43,castin,PRADGO02,PRADGO10}.

This paper is arranged as follows. The scaling formalism applied to the Wigner function and the class of many-body systems for which it is valid are presented in Sec.~\ref{WF}. Section \ref{SCPS} derives the maximum compression factor in space that one can obtain for a given kick force depending on the evolution law of the dilation factor. In Sec.~\ref{SCMS}, we provide the Wigner formalism for a general compression and displacement in momentum space. We also show how a time reversal operator can be applied. Section~\ref{SQC} provides a concrete comparison between non interacting and interacting cases. The issue of anharmonicities is investigated in Sec.~\ref{SAB}. The last section explores more involved phase space manipulations for improving the robustness of compression protocols.

\section{Wigner function for self-similar many-body systems}
\label{WF}

The many-body quantum systems that exhibit self- similar dynamics include the Calogero-Sutherland model \cite{a39}, the Tonks Girardeau gas \cite{a14,a16},
certain Lieb-Liniger states \cite{a40}, Bose-Einstein condensates \cite{a10,a41} even in the presence of dipolar interactions \cite{a42}, strongly interacting gas mixtures \cite{a43}, strongly interacting quantum gases whose collisions are described by the unitary limit \cite{castin}, etc. A non-interacting classical gas described by its phase space distribution function governed by the Boltzmann equation belongs also to the same class of problems \cite{PRADGO02}. 

The formalism that we use relies on the Wigner function $W$ associated with the many-body wavefunction. It is defined via the one-body reduced density matrix $g_1(x,y;t)$:
\begin{equation}
W(x,p;t)=\frac{1}{\pi\hbar}\int g_1(x+y,x-y;t) e^{2ipy/\hbar}{\rm d}y.
\end{equation}
The self-similar dynamics after phase imprinting or in the presence of a constant acceleration field  $g$ involve two time-dependent parameters,  $\alpha(t)$ and $\eta(t)$ \cite{PRADGO14}:
\begin{equation}
g_1(x,y;t)=\frac{1}{\alpha}g_1\left( \frac{x-\eta}{\alpha},\frac{y-\eta}{\alpha};0\right)e^{i(S(x,t)-S(y,t))}, 
\end{equation}
where the time-dependent dilation factor fulfills $\ddot \alpha = \omega_0^2/\alpha^\xi$ (the exponent $\xi$ depends on the specific system that is considered) for a free propagation of a many-body wave function initially at rest in a 1D harmonic confinement of angular frequency $\omega_0$. The time-dependent function $\eta$ accounts for the center of mass motion $\ddot \eta=g$ where $g$ is a constant acceleration field. The phase $S(x,t)$ is given 
by
\begin{equation}
S(x,t)=\frac{m\alpha}{\hbar}\left[  \dot\eta \left( \frac{x-\eta}{\alpha} \right)  + \frac{\dot \alpha}{2 \alpha}\left( \frac{x-\eta}{\alpha} \right)^2\right].
\end{equation}
As a result of the self-similar dynamics, the instantaneous Wigner function $W(x,p;t)$ is simply related to the initial Wigner $W_0(x,p)$ function through the relation
$W(x,p;t)=W_0(X,P)$  with $X=(x-\eta)/\alpha$ and $P=\alpha (p-m\dot\eta)-m\dot\alpha (x-\eta)$ \cite{PRADGO10,PRADGO14}.

Without loss of generality and for pedagogical reasons, we shall compare quite often in the following two specific cases of pure state: (i) a Gaussian wave packet without interactions ($\xi=3$), and (ii) the mean-field wavefunction associated with a Bose-Einstein condensate (BEC) in the Thomas-Fermi regime ($\xi=2$).
In both cases, the initial wavefunction reads $\Psi(x,0)=n_0^{1/2}(x)$, where $n_0$ is the atomic density. In the former case, the Gaussian wavefunction reads $n_0(x)=e^{-x^2/\sigma_0^2}/(\pi^{1/2}\sigma_0)$ and can be considered as the ground state of a harmonic trap of angular frequency $\omega_0$ ($\sigma_0=(\hbar/m\omega_0)^{1/2}$). In the latter case of a Bose-Einstein condensate in the Thomas-Fermi regime, the density in the same harmonic potential reads $n_0(x)= (\mu-m\omega_0^2x^2/2)/\tilde g = m\omega_0^2(R_{\rm TF}^2-x^2)/(2\tilde g)$ where $\mu$ is the chemical potential, $\tilde g$ the strength of the interactions and $R_{\rm TF}= (2\mu/m\omega^2_0)^{1/2}$ the Thomas-Fermi radius.

\section{Compression in position space}
\label{SCPS}
\begin{figure}
\includegraphics[width=8cm,angle=0]{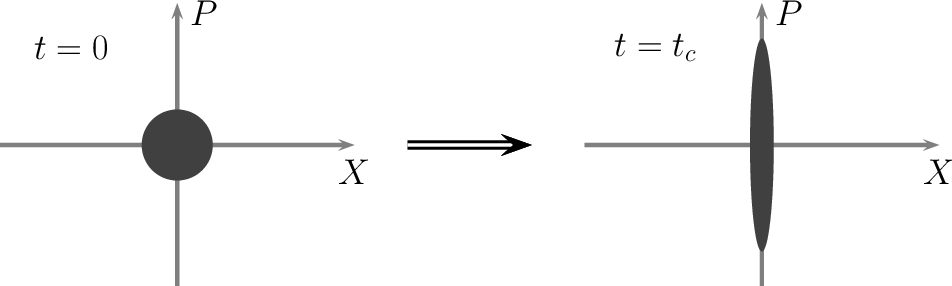}
\caption{Focalisation of a wavepacket by applying suddenly an attractive harmonic force on a very short amount of time.}
\label{figfoca}
\end{figure}

The simplest phase space manipulation is the position squeezing of the wave packet (see Fig.~\ref{figfoca}).
A first strategy consists in increasing the trap strength adiabatically to the desired value. Alternatively, one can use a much faster approach by applying suddenly and for a very short amount of time $\Delta t$ a linear force, $F=-m\omega^2 x$.  This action amounts to setting the time derivative of the dilation factor $\alpha$, $\dot\alpha(0^+)=-\omega^2\Delta t$. Such a force can be realized either with a pulsed magnetic or optical trap. From the differential equation fulfilled by the time-dependent parameter $\alpha$,   
we deduce an energy-like constant of motion:
\begin{equation}
\frac{ {\rm d}}{ {\rm d} t } \left[\left(\frac{ {\rm d}\alpha }{ {\rm d} t }\right)^2 +\omega_0^2\frac{\alpha^{-\xi+1}}{1-\xi} \right]=0.
\label{eqalphaen} 
\end{equation}
This equation is particularly well suited to determine the minimum value of the dilation factor $\alpha_{\rm min}$ resulting from the kick force: 
\begin{equation}
\alpha_{\rm min} (\xi;\omega,\Delta t)= \left( 1 + (1-\xi)\frac{\omega^2}{\omega_0^2}(\omega\Delta t)^2 \right)^{1/(1-\xi)}.
\end{equation}
The wavepacket reaches its minimum size, $(\Delta r)_{\rm c} = \alpha_{\rm min}(\Delta r)_0$, at time $t_{\rm c}\sim 1/\omega_0$ where $(\Delta r)_0$ is its initial size.
By conservation of the phase space volume, the velocity dispersion at that time is increased by the same factor $(\Delta v)_{\rm c} = (\Delta v)_0/\alpha_{\rm min}$.
As intuitively expected, we find $\alpha_{\rm min} (3;\omega,\Delta t)<\alpha_{\rm min} (2;\omega,\Delta t)$. Indeed, the repulsive interactions encapsulated in the exponent $\xi=2$ for a BEC in the Thomas-Fermi regime limit the compressibility of the gas compared to its interaction free counterpart ($\xi=3)$. To ensure that the force remains sudden, one has to fulfill the condition $\omega \Delta t \ll 1$. At its minimum size, the wavefunction exactly coincides with that of the ground state of a harmonic oscillator potential of angular frequency $\omega_c=\omega_0[\alpha_{\rm min} (\xi;\omega,\Delta t)]^{1/(1-\xi)}$ i.e. if we apply suddenly at $t_{\rm c}$ such a harmonic potential, the packet remains unchanged because of the size matching and the phase cancellation, $S(x,t_{c})=0$.

\section{Compression in momentum space}
\label{SCMS}

To reduce the velocity dispersion, one could decrease progressively the trap strength. However the lower the final angular frequency, the larger the time required to ensure the adiabaticity criterion. Furthermore, the sensitivity to low frequency noise in experimental setups limits ultimately the achievable velocity dispersion. To circumvent those limitations we investigate a two-step protocol initially studied for interaction free wave packet and commonly referred to as the Delta-kick cooling protocol \cite{DKick97}.

\begin{figure}
\includegraphics[width=8cm,angle=0]{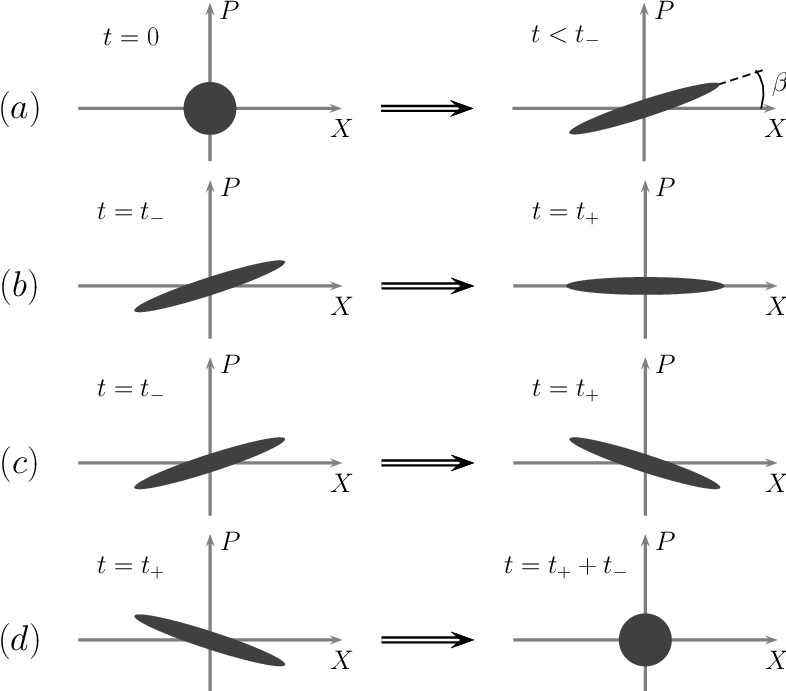}
\caption{To squeeze the wavepacket along the $P$-axis (a), two successive steps are considered: a free expansion (a) followed by a delta-kick force with the right strength to ensure the rotation by the appropriate angle $\beta^*$ (b). A rotation by twice this angle (c) simply obtained by an applied force twice as large as that used for the squeezing, reverses the position-velocity correlation and therefore acts as a time reversal operator. The complete time reversal process implies three steps: free propagation during a finite time $\tau$ (a), rotation by an angle 2$\beta^*$ (c) and a free propagation over a duration equal to $\tau$ (d).}
\label{figtoto1}
\end{figure}

We detail this protocol for a pure state. The generalization to self similar many body wavefunction is straightforward \cite{adolfo}. In such a protocol, the wavefunction first expands freely under a constant acceleration field $g$ during a time interval $t_-$. Figure \ref{figtoto1}(a) represents an example of such an evolution in the case of a free expansion ($g=0$). Then at time $t=t_-$, a force $F=-m\Omega^2(x-x_0)$ is applied suddenly and for a very short amount of time (short with respect to all other timescales), $x_0$ accounts for a position offset between the center of the applied harmonic potential and the center of the atomic cloud. As a result, the wavefunction acquires an extra phase factor just after the application of the force (at time $t_+$ such that $|t_+-t_-|\ll t_-$)
$$
\psi(x,t_+)=\exp \left\{  -i\gamma (x-x_0)^2 \right\} \psi(x,t_-),
$$
where $\gamma=m\Omega^2(t_+-t_-)/(2\hbar)$. Applying this force amounts to rotating the phase space volume (see Fig.~\ref{figtoto1}(b)). The corresponding Wigner function reads
$$
W(x,p,t_+)=\frac{1}{N\pi\hbar}\int {\rm d}Y f_0(X,Y)e^{2iPY/\hbar}
$$
with $P = \alpha\left(p + 2\hbar \gamma (x-x_0)-m\dot\eta - m\dot\alpha X \right)$, 
$f_0(X,Y)=n_0^{1/2}(X+Y)n_0^{1/2}(X-Y)$ and $Y=y/\alpha$. From this expression, we infer 
the mean value of the momentum
\begin{eqnarray}
\langle p \rangle &  =  & \int {\rm d}x{\rm d}p\, p\, W(x,p,t_+) \nonumber \\
&  =  & \frac{1}{N\pi\hbar}\int {\rm d}X{\rm d}P \left[ \frac{P}{\alpha} + aX + p_t  \right]f_0(X,Y)e^{2iPY/\hbar}  \nonumber \\
&  =  & \frac{\langle p \rangle_0}{\alpha} + a\langle x \rangle_0 + p_t
\end{eqnarray}
with $a=m\dot\alpha - 2\hbar\gamma\alpha$ and 
\begin{equation}
p_t=m \dot\eta-2\hbar\gamma(\eta-x_0).
\label{pt}
\end{equation}
The quantity $p_t$ provides an offset in momentum space.
A similar calculation for $\langle p^2 \rangle$ yields
$$
\langle p^2 \rangle = \frac{\langle p^2 \rangle_0}{\alpha^2} + a^2\langle x^2 \rangle_0 + p_t^2 + \frac{2p_t}{\alpha}\langle p \rangle_0 +2ap_t\langle x \rangle_0 + \frac{2a}{\alpha}\langle xp \rangle_0.
$$
We conclude that after the two-step protocol, the variance of the momentum reads
$$
(\Delta p)^2 = \frac{(\Delta p)_0^2}{\alpha^2} + a^2(\Delta x)_0^2 + \frac{2a}{\alpha}(\langle xp \rangle_0-\langle x \rangle_0\langle p \rangle_0).
$$

\begin{figure}
\includegraphics[width=8cm,angle=0]{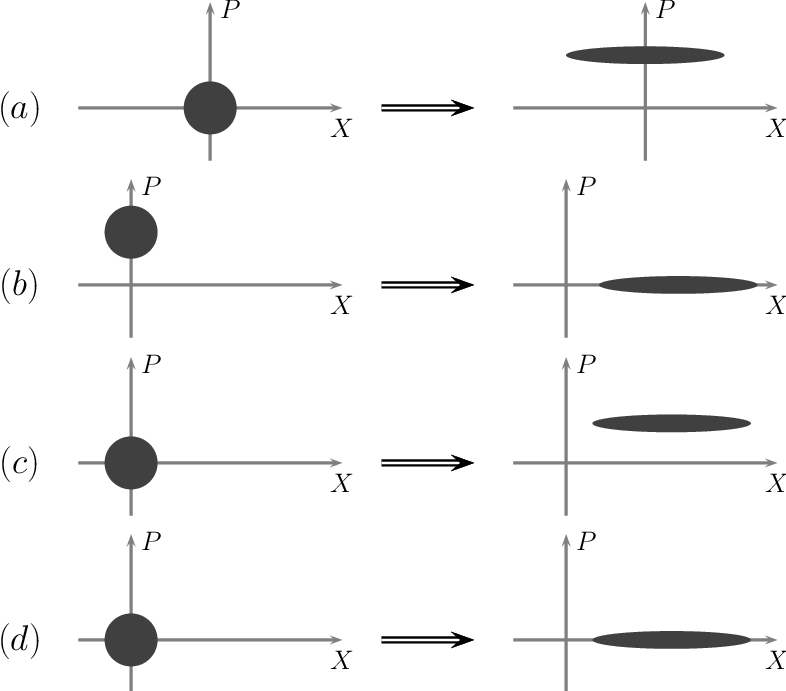}
\caption{(a) and (b): Using a linear kick force with an offset, the cooling and translation on the momentum axis can be obtained simultaneously. (c) and (d) The final position of the squeezed wavepacket can be chosen in phase space by an appropriate sequence of evolution under a constant force followed by the application of a sudden force with or without offset.}
\label{figtoto2}
\end{figure}
 
The optimal choice to minimize the momentum dispersion is $a=0$, i.e. $m\dot\alpha=2\hbar\gamma\alpha$. It corresponds to the optimal rotation angle $\beta=\beta^*$ illustrated in Fig.~\ref{figtoto1}(b). The value to be used for $\gamma$ therefore depends on the evolution of the dilation factor $\alpha$ and its strength is tunable experimentally by the control of the quantity $\Omega^2(t_+-t_-)$. The corresponding velocity dispersion reduction is $\Delta v(t_+) = \Delta v(0)/\alpha_-$ with $\alpha_-=\alpha(t_-)$. This result is particularly interesting since it yields a reduction of the velocity dispersion with respect to the \emph{initial} one in a short amount of time, and therefore shortcuts any adiabatic transformation \cite{STA2013}. In the absence of interactions, we find $\alpha(t)=(1+\omega_0^2t^2)^{1/2}$ yielding $\Delta v(t) =\Delta v(0)/(1+\omega_0^2t^2)^{1/2}$. The specific angle value, $\beta=\beta^*$,  is also the one that cancels the quadratic term in $x$ in the phase $S(x,t_+)$. As a result, the wavefunction is at $t_+$ in the form $\Psi(x,t_+) = \sqrt{n_0( x/\alpha_-)}/\alpha_-^{1/2}\exp(i\kappa x)$ with a time-dependent global phase factor. 

The linear term in $x$ in the phase accounts for the mean momentum $\hbar\kappa$ that can be acquired in the two-step protocol and/or depends on the initial conditions. For the optimal choice $\beta=\beta^*$, the mean momentum is equal to $\langle p \rangle=\langle p \rangle_0/\alpha + p_t$. After the two-step protocol, the momentum is therefore translated by a quantity $\langle p \rangle-\langle p \rangle_0=\langle p \rangle_0(\alpha_-^{-1}-1)+p_t$. According to Eq.~(\ref{pt}), there are two independent ways to communicate a mean momentum to the packet while reducing the momentum dispersion: either using a constant acceleration field ($g\neq 0$), or applying an off-center sudden harmonic potential ($x_0\neq 0$). These two possibilities can also be combined (see Eq.~(\ref{pt})). In Fig.~\ref{figtoto2}, a few examples of possible phase space manipulations are described.
Using for the phase imprinting a harmonic potential with an offset ($x_0\neq 0$), one can simultaneously launch a packet and cool it (see Fig.~\ref{figtoto2}(a)). Let us give an order of magnitude with realistic values for a non interacting wave packet with $\omega_0=2\pi\times 150$ Hz and a free expansion time of 5 ms, we find $\dot \alpha / \alpha = 192$  $s^{-1}$, for an offset of $x_0=20$ $\mu$m, we get $\bar{v} =3.8$ mm/s.

The application of a sudden force imprints a quadratic with position phase and is thus in close analogy with the action in Fourier space of a lens in optics. The role of the standard delta-kick cooling protocol is to remove the phase acquired during the propagation. The velocity resulting from the off-center delta-kick cooling corresponds to the use of two lenses with opposite focal length  $f$ but with a displacement of their center by a quantity $a$. the optical analog is an optical beam that would arrive from infinity would be deflected by an angle $-a/f$. This angle plays the same role as the offset mean velocity. 

We have already seen that the choice of $\beta$ that provides the optimal rotation amounts to cancelling the quadratic term of the phase of the wavefunction.
In the absence of center of mass motion, if one rotates by twice this angle $\beta=2\beta^*$, the phase is reversed: $S(x,t_+)=-S(x,t_-)$. As a result, $\Psi(x,t_+)=\Psi^*(x,t_-)$, an operation that corresponds exactly to the time reversal operator. This means that the wavefunction will reconcentrate towards its original form despite the repulsive interactions if any. This latter situation is illustrated in Fig.~\ref{figtoto1}(c).
The free evolution is then a refocussing towards the initial state (see Fig.~\ref{figtoto1}(d)). Actually, this time reversal technique enables one to determine precisely the gain on velocity reduction. Indeed, the value may be so low that the standard time-of-flight technique can no more be used in standard setups. In this case, the time reversal operation provides a way to infer the velocity dispersion obtained for a rotation by the angle $\beta^*$ in the delta-kick cooling protocol. We shall see another application of this time reversal possibility in Sec.~\ref{spinecho}.

\section{Quantitative comparison between the non-interacting and interacting case}
\label{SQC}
 
The initial velocity dispersion for a trapped Bose-Einstein condensate in the Thomas-Fermi regime is small compared to its non-interacting counterpart since the size of the atomic sample is larger. However, the velocity dispersion increases when the confinement is removed due to the conversion of the interaction energy into kinetic energy. 

In the following, we propose a quantitative comparison of the relative performances of the two-step protocol for the interaction free Gaussian wavefunction and the wavefunction of a BEC in the Thomas-Fermi (TF) regime. For the sake of simplicity, we restrict ourselves to the situation in which the wave packet is initially at rest and experiences no constant acceleration field ($g=0$). The generalization to $g\neq 0$ of the solution worked out hereafter is straightforward.

\begin{figure}
\includegraphics[width=8cm,angle=0]{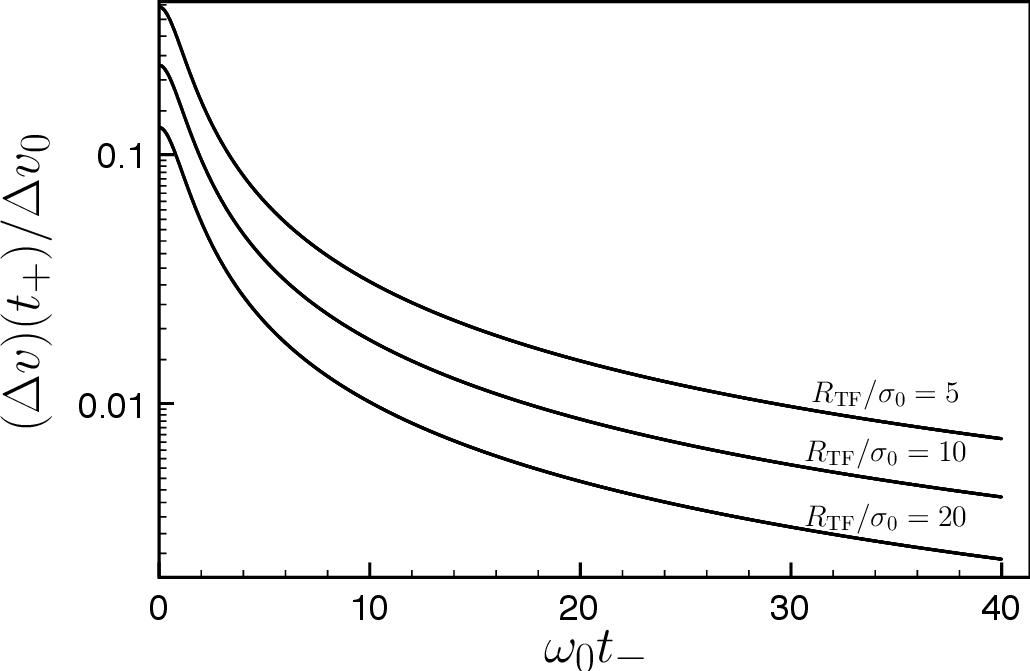}
\caption{Velocity dispersion just after the delta-kick cooling protocol as a function of the time, $t_-$,  at which the rotation is performed and for various value of the ratio between the initial Thomas-Fermi radius $R_{\rm TF}$ and the harmonic length $\sigma_0$. The velocity dispersion is normalized to that of the Gaussian ground state wavefunction in the same initial harmonic confinement, and the time is normalized to $\omega_0^{-1}$.}
\label{fig2}
\end{figure}

Let us first recall the expression for the velocity dispersion of a BEC in the TF regime \cite{Vermersch11}:
\begin{equation}
(\Delta v_{\rm TF})(0) = \Delta v_0  \frac{\sqrt{2}}{\delta} \sqrt{{\rm ln}\left(  \theta \delta\right)},
\end{equation}
with $\theta\simeq 1.373475...$ a numerical factor, $\delta =R_{\rm TF}/\sigma_0$ and $\Delta v_0=(\hbar\omega_0/2m)^{1/2}$ the velocity
dispersion for the non-interacting Gaussian wavefunction. 
The optimal rotation angle $\beta^*$ defined by $a=0$ takes into account the deformation of the phase space surface induced by the interactions through the exponent $\xi$ in the evolution of the dilation factor $\alpha$. 
For $t<t_-$ the velocity dispersion is given by \cite{PRADGO14}
\begin{equation}
(\Delta v)(t)=\Delta v_{\rm TF}(0) \left[  \frac{1}{\alpha(t)^2} + \left(1-\frac{1}{\alpha(t)}\right)\frac{\delta^2}{ {\rm ln}\left(  \theta \delta\right)} \right]^{1/2}. 
\end{equation}
 At $t=t_+$, $\Delta v(t_+)=\Delta v_{\rm TF}(0)/\alpha_-$. 
 Figure \ref{fig2} gives the evolution of the ratio $(\Delta v)(t_+)/(\Delta v_0)$ as a function of time for different values of the ratio $\delta=R_{\rm TF}/\sigma_0$ just after applying the two-step protocol. At first sight, one could conclude that there is a very impressive gain in terms of reduction of the velocity dispersion in applying the Delta-kick cooling technique on a many-body wavefunction.

However, a more rigorous comparison requires the study of the evolution after the two-step protocol i.e. $t>t_+$. Indeed the interaction energy remains as a reservoir that can significantly increase the velocity dispersion.  From an analytical perspective, the answer turns out to be not that simple despite the fact that we know exactly the expression for the wavefunction at $t_+$: $\Psi(x,t_+) = \sqrt{\tilde n(x)}$ where $\tilde n(x)=n_0( x/\alpha_-)/\alpha_-^{1/2}$ with $\alpha_-=\alpha(t_-)$. Indeed, this inverse parabola shape corresponds to that of a Bose-Einstein condensate in the Thomas-Fermi regime for which the interactions strength would be $g_+=\tilde g/\alpha_-^2$ and the confinement strength would also be reduced $\omega_+=\omega_0/\alpha_-$. The 1D dimensionless $\chi$ parameter given by the ratio of the interaction energy over the kinetic energy, is thereby drastically reduced $\chi_+=\chi / \alpha_-^{5/2}$. For a standard expansion (by a factor 3-10), this reduction factor is large and the wavefunction cannot be any longer considered as the one of a Bose-Einstein condensate in the Thomas-Fermi regime. The dilution is too important, and the interaction energy is not important anymore in comparison to the kinetic energy. As a result, a scaling ansatz can no more be used to account for the time evolution of the wavefunction at $t>t_+$; only numerics can provide the exact evolution \cite{peS}. Alternatively, one can analytically obtain the asymptotic value of the velocity dispersion. For this purpose, we need to calculate the kinetic energy $E_{\rm kin}(t_+)$ and the interaction energy $E_{\rm int}(t_+)$ and we shall use the conservation of energy:
\begin{equation}
(\Delta v)^2(\infty) =  \frac{2}{mN} \left(  E_{\rm kin}(t_+) + E_{\rm int}(t_+)\right).
\label{encons}
\end{equation}
We can calculate directly the interaction energy: $E_{\rm int}(t_+) = E_{\rm int}(0)/\alpha_-$ and 
$E_{\rm int}(0) = N\hbar\omega_0 \delta^2/5$. The expression for the kinetic energy can be obtained from Ref.~\cite{PRADGO14}
$$
E_{\rm kin}(t_+) = \frac{N\hbar \omega_0}{2\alpha_-^2\delta^2} {\rm ln}\left(  \theta \delta\right).
$$

From Eq.~(\ref{encons}), we infer the asymptotic velocity dispersion for $t \gg t_+$ 
$$(\Delta v)(\infty) =  \Delta v_0 \left[   \frac{4\delta^2}{5\alpha_-} + \frac{2{\rm ln}\left(  \theta \delta\right)}{\alpha^2_-\delta^2} \right]^{1/2}.$$

\begin{figure}
\includegraphics[width=8cm,angle=0]{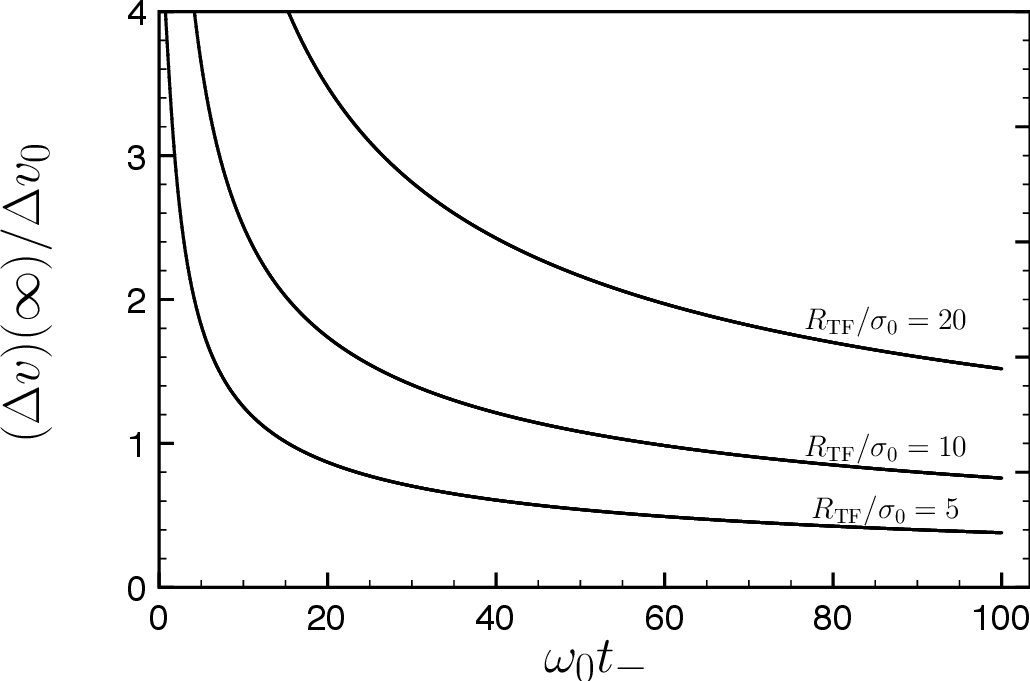}
\caption{Asymptotic velocity dispersion as a function of the time, $t_-$, at which the rotation is performed and for various value of the ratio between the initial Thomas-Fermi radius $R_{\rm TF}$ and the harmonic length $\sigma_0$. Same normalization as in Fig.~\ref{fig2}.}
\label{fig3}
\end{figure}

In Fig.~\ref{fig3}, the evolution of the ratio $\Delta v(\infty)/\Delta v_0$ as a function of time for different values of the ratio $\delta=R_{\rm TF}/\sigma_0$ is plotted. We obtain a much more moderate gain. Interestingly, this ratio always goes below one if the phase imprinting is applied after a sufficiently long free expansion time, meaning that the contribution of repulsive interactions to the velocity dispersion can be completely washed out using the two-step protocol and that a dispersion even below the non-interacting case can be achieved. It is worth noticing that this latter prediction is the most pessimistic one. Indeed the time required to reach this asymptotic value increases considerably (by at least a factor $\alpha_-$) after the phase space rotation because of the dilution. For an experiment carried out on a short time scale compared to $\alpha_-/\omega_0$, one can really benefit from the gain in velocity dispersion presented in Fig.~\ref{fig2}.

\section{Aberration}
\label{SAB}

In this section, we investigate the influence of anharmonicities in the suddenly applied potential for the delta-kick cooling technique. Calculations are carried out for a pure state initially at rest ($\eta=0$). We consider that the phase imprinting does not correspond exactly to a quadratic force but contains an extra cubic correction. We propose a perturbative analysis of this effect that would correspond to spherical aberration in optics. Under our assumption, the wavefunction after the phase imprinting reads
\begin{eqnarray}
\psi(x,t_+) & = & e^{-i\gamma x^2 + i\epsilon x^3}\psi(x,t_-) \nonumber \\ & \simeq &  e^{-i\gamma x^2 }\psi(x,t_-) \left( 1 + i\epsilon x^3 - \frac{\epsilon^2x^6}{2}\right).
\end{eqnarray}
The Wigner function contains an extra factor $\Lambda(x,y)$ compared to the case without anharmonicitities
\begin{eqnarray}
&& W(x,p,t_+)  =  \frac{1}{N\pi \hbar} \int \psi^*(x+y,t_+) \psi(x-y,t_+)e^{2ipy/\hbar}dy \nonumber \\
& & =  \frac{1}{N\pi \hbar} \int \psi^*(x+y,t_-) \psi(x-y,t_-)\Lambda(x,y)e^{2i\tilde{p}y/\hbar}dy
\end{eqnarray}
with $\tilde{p}=p+2\hbar\gamma x$. The expansion up to the second order in $\epsilon$ of $\Lambda$ reads
$\Lambda(x,y) \simeq  1 -2i\epsilon y(3x^2+y^2) -2\epsilon^2y^2 \left( 9x^4 + 6x^2y^2 + y^4  \right).$ We deduce
$$
W(x,p,t_+)  =  \frac{1}{N\pi \hbar} \int dY g_0(X,Y) e^{2iPy/\hbar}dy \nonumber \\
$$
with $g_0(X,Y)=f_0(X,Y)\Lambda(\alpha X,\alpha Y)$, $P=\alpha \tilde{p}-m\dot\alpha x$, $X=x/\alpha$, $Y=y/\alpha$. Using the same notation as previously, we get

\begin{eqnarray}
\langle p \rangle &  =  & \int {\rm d} x {\rm d} p\, p\, W(x,p,t_+) \nonumber \\
&  =  & \frac{1}{N\pi\hbar}\int dXdP \left[ \frac{P}{\alpha} + aX   \right]g_0(X,Y)e^{2iPY/\hbar}  \nonumber \\
&  =  &  \frac{1}{N\pi\hbar }  \frac{1}{\alpha}\int dX \left( -\frac{\pi \hbar^2}{2i}\right) \partial_Yg_0(X,Y)\bigg|_{Y=0}
+ 0 \nonumber \\
&  =  & \frac{1}{N\pi\hbar }  \frac{1}{\alpha}  \left( -\frac{\pi \hbar^2}{2i}\right) (-6i\epsilon \alpha^3) \int X^2\psi_0(X)dX  \nonumber \\
&  =  & 3\epsilon \alpha^2\hbar \langle x^2 \rangle_0.
\end{eqnarray}
A similar calculation for $\langle p^2 \rangle$ yields
$\langle p^2 \rangle = \langle p^2 \rangle_0/\alpha^2+ a^2 \langle x^2 \rangle_0 + 9\epsilon^2 \alpha^4\hbar^2 \langle x^4 \rangle_0.$
We conclude that the optimal variance of the momentum is given by
$$
( \Delta p)^2 = ( \Delta p)_0^2/\alpha^2 + (\Delta p)_\epsilon^2
$$
where $(\Delta p)_\epsilon^2=9\epsilon^2 \alpha^4\hbar^2 (\langle x^4 \rangle_0-\langle x^2 \rangle_0^2)$.
We conclude that the optimal angle remains the same as in the case without aberration but the reduction of the velocity dispersion is clearly affected by the nonlinearity.
 The effect is particularly strong in the regime in which the two terms are on the same order of magnitude i.e. for $t>t_\epsilon$ with $t_\epsilon$ defined by $\alpha(t_\epsilon)=(\Delta p)_0/(\Delta p)_\epsilon$.

\section{Higher order protocols}
\label{spinecho}

\begin{figure}
\includegraphics[width=8cm,angle=0]{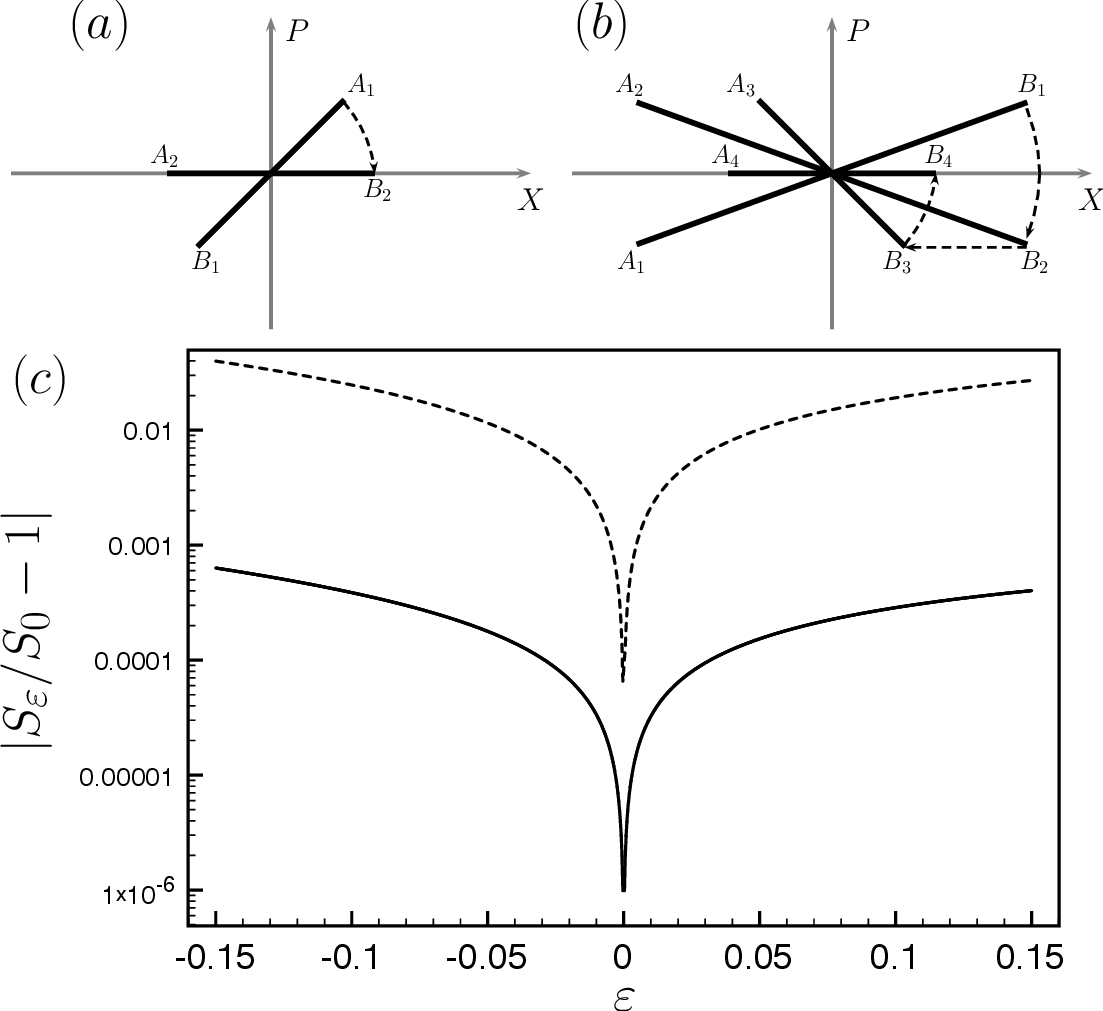}
\caption{(a) Strategy 1: simple phase space rotation (dashed line). (b) Strategy 2: spin-echo like phase space manipulation (solid line). For the sake of simplicity, we only represent the long axis of the ellipse shape of the phase space volume. (c) The relative error on the phase $|S_\varepsilon/S_0-1|$ as a function of the frequency mistmatch $\varepsilon$.}
\label{figspinecho}
\end{figure}

In this section, we design an ultra robust many step protocol inspired by the spin-echo technique well-known in NMR \cite{bookae}. This technique used for robust quantum information processing allows for an improved fidelity of a $\pi-$pulse operation in the presence of a dispersion of Rabi frequencies. Its simplest version relies on a sequence of three successive pulses: $(\pi/2)_Y(\pi)_X(\pi/2)_Y$. After the first pulse the spin directions are spread over a finite angle about $\pi/2$. The spins with the largest Rabi frequency have rotated by more than $\pi/2$. The second pulse reverses on the Bloch sphere the relative position of the slowest and largest Rabi frequency spins. Finally the last pulse refocuses all spins. 

The transposition of this technique in phase space manipulation is summarized in Fig.~\ref{figspinecho}. We compare the robustness against fluctuations of the initial trap frequency ($\omega=\omega_0(1+\varepsilon)$) of two different schemes aiming at reducing the velocity dispersion. This provides a way to probe the robustness against anharmonicities. The first one is the standard delta-kick cooling strategy (see Fig.~\ref{figspinecho}(a)), it consists of a free propagation over a time $T$ followed by a sudden clockwise optimal phase space rotation. The second one involves a free propagation over a time $T'$, a sudden time reversal force pulse, a free propagation over a time $T''$ and a sudden anti-clockwise rotation (see Fig.~\ref{figspinecho}(b)). The time interval $T'$ is chosen so that $\alpha(T')=2\alpha(T)$ and the time $T''$ is chosen so that the final state is the same as in the first strategy. The time reversal pulse plays the same role as the $(\pi)_X$ pulse in the spin-echo protocol, it reverses the position of the slowest and fastest atoms i.e. the atoms that experience initially a lower or larger trap frequency. The final anti-clockwise pulse refocuses the different trajectories.
 The relative error on the phase $|S_\varepsilon/S_0-1|$ as a function of the frequency mismatch $\varepsilon$ can be appreciated in Fig.~\ref{figspinecho}(c). The direct delta-kick strategy yields a relative error that scales as $\varepsilon^2$ while $\rho \sim \varepsilon^4$ for the spin-echo like strategy.  This latter strategy turns out to be much more robust.

In conclusion, we have developed a general framework to calculate quantitatively a wide variety of phase space manipulations. Calculations were essentially carried out for a pure state but remain valid for a large class of many-body systems that exhibit self-similar dynamics \cite{adolfo}. 
We have discussed how a substantial gain on the velocity dispersion reduction can be obtained even when the interaction energy is initially very large in comparison to the kinetic energy. These results are potentially important for the initial state preparation in many different types of cold atom experiments.

We are grateful to A. Gauguet, F. Damon and J. G. Muga for useful comments.
This work was supported by Programme Investissements d'Avenir under the program ANR-11-IDEX-0002-02, reference ANR-10-LABX-0037-NEXT and the Institut Universitaire de France.

\end{document}